\documentclass[conference,a4paper]{IEEEtran}

\usepackage[T1]{fontenc}
\usepackage[utf8]{inputenc}
\usepackage[english]{babel}
\usepackage{cite}
\usepackage{amsmath,amssymb,amsfonts}
\usepackage[ruled, lined, linesnumbered, commentsnumbered, longend]{algorithm2e}
\usepackage{algorithmic}
\usepackage{graphicx,color}
\usepackage{textcomp}
\usepackage{soul}
\usepackage{enumerate}
\usepackage{comment}
\usepackage{multirow}
\usepackage{tabu}
\usepackage{float}
\usepackage[autostyle]{csquotes}
\usepackage[utf8]{inputenc}
\usepackage{mathtools}
\usepackage[autostyle]{csquotes}
\usepackage{textcase}
\usepackage{float}
\usepackage{tabu}
\usepackage{multirow}
\usepackage[ruled, lined, linesnumbered, commentsnumbered, longend]{algorithm2e}
\usepackage{bbm}
\usepackage{siunitx}
\usepackage{booktabs}
\usepackage{textcomp}
\usepackage{anyfontsize}
\usepackage{url}
\usepackage{subcaption}

\usepackage[acronym,toc,nomain]{glossaries} 
\makeglossaries
\newacronym{mapc}{MAPC}{multi-access point coordination}
\newacronym{ap}{AP}{access point}
\newacronym{ai}{AI}{artificial intelligence}
\newacronym{obss}{OBSS}{overlapping basic service set}
\newacronym{uhr}{UHR}{Ultra High Reliability}
\newacronym{tg}{TG}{Task Group}
\newacronym{drl}{DRL}{deep reinforcement learning}
\newacronym{rl}{RL}{reinforcement learning}
\newacronym{ppo}{PPO}{proximal policy optimization}
\newacronym{ml}{ML}{machine learning}
\newacronym{qos}{QoS}{quality of service}
\newacronym{csr}{Co-SR}{coordinated spatial reuse}
\newacronym{sr}{SR}{spatial reuse}
\newacronym{mnp}{MNP}{maximum number of packets}
\newacronym{op}{OP}{oldest packet}
\newacronym{tat}{TAT}{traffic-alignment tracker}
\newacronym{mdp}{MDP}{Markov decision process}
\newacronym{mlp}{MLP}{multi-layer perceptron}
\newacronym{sinr}{SINR}{signal-to-interference-plus-noise ratio}
\newacronym{snr}{SNR}{signal plus noise ratio}
\newacronym{per}{PER}{packet error rate}
\newacronym{mcs}{MCS}{modulation and coding scheme}
\newacronym{dcf}{DCF}{distributed coordination function}
\newacronym{mapc-icf}{MAPC-ICF}{MAPC-initial control frame}
\newacronym{mapc-icr}{MAPC-ICR}{MAPC-initial control response}
\newacronym[longplural={transmission opportunities}]{txop}{TXOP}{transmission opportunity}
\newacronym{sifs}{SIFS}{short interframe space}
\newacronym{difs}{DIFS}{DCF interframe space}
\newacronym{cw}{CW}{contention window}
\newacronym{mapc-tf}{MAPC-TF}{MAPC-trigger frame}
\newacronym{mpdu}{MPDU}{MAC protocol data unit}
\newacronym{ampdu}{A-MPDU}{aggregate MAC protocol data unit}
\newacronym{back}{BACK}{block acknowledgment}
\newacronym{sta}{STA}{station}
\newacronym{bss}{BSS}{basic service set}
\newacronym{rfa}{RFA}{radio frequency} 
\newacronym{los}{LoS}{line of sight} 
\newacronym{mab}{MAB}{multi-armed bandit} 
\newacronym{ma-mab}{MA-MAB}{multi-agent multi-armed bandit} 
\newacronym{ucb}{UCB}{upper confidence bound} 
\newacronym{ofdm}{OFDM}{orthogonal frequency-division multiplexing}
\newacronym{ofdma}{OFDMA}{orthogonal frequency-division multiple access}
\newacronym{fifo}{FIFO}{first in first out}
\newacronym{hol}{HoL}{head-of-line}
\newacronym{edca}{EDCA}{enhanced distributed channel access}
\newacronym{nav}{NAV}{network allocation vector}

\definecolor{boristext}{rgb}{0.3, 0.36, 0.88}
\definecolor{boriscomments}{rgb}{0.83, 0.0, 0.0}

\definecolor{davidcomments}{rgb}{0.0, 0.0, 0.83}

\begin{document}


\title{Empowering Multi-Access Point Coordination through Machine Learning-based Scheduling}
\title{Deep Reinforcement Learning-Based Scheduling for IEEE 802.11bn Multi-Access Point Coordination}
\title{Deep Reinforcement Learning-Based Scheduling for Wi-Fi Multi-Access Point Coordination}

\author{
\IEEEauthorblockN{David Nunez$^{\star}$, Francesc Wilhelmi$^{\star}$, Maksymilian Wojnar$^{\sharp}$, Katarzyna Kosek-Szott$^{\sharp}$, \\ Szymon Szott$^{\sharp}$ and Boris Bellalta$^{\star}$, \vspace{0.1cm}
}
\IEEEauthorblockA{$^{\star}$\emph{Wireless Networking, Universitat Pompeu Fabra, Barcelona, Spain}}
\IEEEauthorblockA{$^{\sharp}$\emph{AGH University of Krakow, Poland}}
\IEEEauthorblockN{\thanks{Corresponding author: \emph{david.nunez@upf.edu}.}}
}

\maketitle


\begin{abstract}
Multi-access point coordination (MAPC) is a key feature of IEEE 802.11bn, with a potential impact on future Wi-Fi networks. MAPC enables joint scheduling decisions across multiple access points (APs) to improve throughput, latency, and reliability in dense Wi-Fi deployments. However, implementing efficient scheduling policies under diverse traffic and interference conditions in overlapping basic service sets (OBSSs) remains a complex task. This paper presents a method to minimize the network-wide worst-case latency by formulating MAPC scheduling as a sequential decision-making problem and proposing a deep reinforcement learning (DRL) mechanism to minimize worst-case delays in OBSS deployments. Specifically, we train a DRL agent using proximal policy optimization (PPO) within an 802.11bn-compatible Gymnasium environment. This environment provides observations of queue states, delay metrics, and channel conditions, enabling the agent to schedule multiple AP-station pairs to transmit simultaneously by leveraging spatial reuse (SR) groups. Simulations demonstrate that our proposed solution outperforms state-of-the-art heuristic strategies across a wide range of network loads and traffic patterns. The trained machine learning (ML) models consistently achieve lower 99th-percentile delays, showing up to a 30\% improvement over the best baseline.
\end{abstract}


\maketitle

\section{Introduction}

The rapid growth of wireless traffic, the rise of latency-sensitive applications, and the increasing density of user devices are imposing unprecedented demands on Wi-Fi networks. Traditional contention-based channel access mechanisms do not scale effectively in dense deployments with diverse traffic patterns and high levels of co-channel interference~\cite{costa2015limitations}. These limitations lead to poor~\gls{qos}, large delays, and inefficient spectrum utilization. Future Wi-Fi technologies must embrace more deterministic channel access and latency-sensitive scheduling to meet the demands of emerging use cases such as augmented reality, industrial automation, and dense public venues~\cite{galati2024will}.

In this context,~\gls{mapc}, introduced in IEEE 802.11bn, represents a fundamental architectural shift from uncoordinated to coordinated spectrum sharing. By enabling centralized scheduling decisions across multiple~\glspl{ap},~\gls{mapc} facilitates joint optimization of time, frequency, and spatial resources. 
\Gls{csr} stems as a specific~\gls{mapc} technique to allow multiple~\gls{ap}–\gls{sta}\footnote{In the IEEE 802.11 standard, client devices are referred to as non-~\gls{ap}~\glspl{sta}, which we refer to as~\glspl{sta} in the remainder of this paper.} pairs to transmit concurrently by coordinating~\gls{sr} opportunities across the network~\cite{mypaperCSRAnalyticalModel}. Unlike time---or frequency---division coordination~\cite{val2025wi},~\gls{csr} aims to maximize spectral efficiency by allowing overlapping transmissions whereby interference constraints are satisfied. Among the various~\gls{mapc} strategies,~\gls{csr} offers a compelling trade-off between complexity and performance, making it particularly well-suited for latency-sensitive applications in high-density scenarios while keeping implementation costs low. For this reason, our analysis focuses on~\gls{mapc} networks employing~\gls{csr} as the underlying coordination mechanism. However, the effectiveness of~\gls{mapc} networks using~\gls{csr} hinges not only on the feasibility and quality of parallel transmissions but also on the ability to make timely and informed scheduling decisions under diverse traffic and channel conditions. Designing efficient~\gls{mapc} schedulers remains a challenging task. The scheduler should make decisions influenced by traffic, signal quality, and interference for various independent~\glspl{bss}, resulting in complex relationships that are hard to capture by heuristic algorithms, such as the use of~\gls{mnp},~\gls{op}, and~\gls{tat} scheduling strategies for~\gls{csr}~\cite{my_paper_joint_scheduling}.

In this work, we address the~\gls{mapc} scheduling problem by formulating it as a sequential decision-making task and proposing a model capable of learning scheduling policies from real-time network observations. The proposed solution consists of a~\gls{drl} agent that interacts with an environment through a standardized Gymnasium interface.\footnote{\url{https://gymnasium.farama.org/}} Our main aim lies in demonstrating that the learned policies consistently outperform heuristic-based schedulers, such as those previously proposed in~\cite{my_paper_joint_scheduling}, by effectively adapting to diverse topologies, loads, and traffic patterns. Our results show that the proposed model achieves up to a~30\% reduction in worst-case delay compared to the best-performing non-ML baseline. The main contributions of this paper are summarized as follows:
\begin{itemize}
    \item We formulate the 802.11bn~\gls{mapc} scheduling problem as a sequential decision-making optimization problem.
    \item We propose a~\gls{drl} solution to learn dynamic~\gls{mapc} scheduling policies based on real-time metrics such as the queueing delays of the entire~\gls{obss}.
    \item We develop a novel Gymnasium-compatible simulation environment that integrates AI-driven decision-making with a detailed 802.11bn-compliant simulator, enabling the training and evaluation of DRL-based MAPC schedulers under diverse traffic and channel conditions.
    \item We showcase the performance of our proposed~\gls{drl}-based scheduling solution and compare it to  heuristic baselines such as~\gls{mnp},~\gls{op}, and~\gls{tat}, in a comprehensive set of simulation scenarios.  
\end{itemize}

The rest of this paper is structured as follows. Section~\ref{sec:related_work} delves into the state-of-the-art solutions on~\gls{mapc} and~\gls{csr}. Section~\ref{sec:mapc} discusses the IEEE 802.11bn~\gls{mapc} setup and our scheme for creating~\gls{sr}-compatible groups of devices from multiple~\glspl{bss}. Then, Section~\ref{sec:ml} describes the proposed~\gls{ml} setup and~\gls{drl}-based scheduling solution. The evaluation setup and simulation details are provided in Section~\ref{Sec:evaluation_setup}, whereas the simulation results are presented in Section~\ref{sec:performance}. Section~\ref{sec:conclusions} concludes the paper.


\section{Related Work}
\label{sec:related_work}

Since its introduction by the IEEE 802.11ax standard,~\gls{sr} has received a lot of attention, and many works have attempted to improve its performance by adopting~\gls{ml} techniques. Examples include~\cite{wilhelmi2019collaborative, wilhelmi2019potential, bardou2021improving, huang2022reinforcement}, where~\gls{mab} solutions drive the selection of~\gls{sr} parameters. \glspl{mab} are a suitable framework for online decision-making due to their simplicity and effectiveness, thus being a good match for \mbox{Wi-Fi}'s MAC optimization.

Learning-based approaches have also been explored to tackle the inherent complexity and the combinatorial action space of~\gls{mapc} networks. In~\cite{wilhelmi2024machine}, the authors showcase the potential of~\gls{rl}-based solutions, including decentralized and coordinated approaches, for driving the optimization of~\gls{sr} in dense Wi-Fi deployments. Later, the authors propose a~\gls{ma-mab} approach to jointly configure~\gls{sr} parameters---specifically packet detection thresholds and transmit power---across multiple coexisting~\glspl{ap}, based on coordinated (shared) reward strategies~\cite{Francesc_MABs}. Their results show significant gains in throughput and fairness compared to uncoordinated baselines. Similarly,~\cite{Wojnar_MABS} introduces a hierarchical~\gls{mab} framework for~\gls{csr} group selection in IEEE 802.11bn. Using \gls{rl} to select compatible~\gls{ap} subsets, the study finds that~\gls{ucb}-based bandits offer fast convergence and sustained performance across heterogeneous topologies. Finally, the work in~\cite{maks_CoSR_journal} addresses the~\gls{csr} scheduling problem in IEEE 802.11bn by proposing a practical solution based on~\glspl{mab}, including both flat and hierarchical variants. Simulations and testbed experiments show that hierarchical~\glspl{mab} can improve aggregate throughput by up to 80\% over legacy IEEE 802.11, without compromising the number of assigned per-\gls{sta} transmission opportunities.

Regarding Wi-Fi scheduling, the main focus of this paper, recent efforts have been put into ML as well. In~\cite{li2025reinwifi}, an~\gls{rl}-based scheduling framework is proposed and implemented to optimize the application-layer~\gls{qos} of a Wi-Fi network with commercial devices. The method adjusts contention window sizes and throughput limits using a Q-network trained from historical \gls{qos} feedback, achieving significant improvements over \gls{edca} in a real testbed. In addition, the authors in~\cite{Cui_2019} use deep learning to schedule transmissions based solely on node location, bypassing channel estimation and handling dense interference scenarios. Linked closely to this paper,~\cite{du2025deep} proposes a~\gls{drl}-based solution to optimize~\gls{sr} in 802.11bn networks. However, the approach considered in~\cite{du2025deep} is substantially different from ours, as~\gls{drl} decisions are used to perform channel access and link adaptation. In this paper, instead, we focus on the selection of suitable~\gls{sr} groups for transmission, as part of a~\gls{drl}-based~\gls{mapc} scheduler. This is something that has already been tackled in~\cite{my_paper_joint_scheduling}, where non-\gls{ai} schedulers are evaluated. Our paper advances the state-of-the-art by demonstrating the potential of~\gls{ml} solutions, which are expected to effectively address the increasing complexity in multi-\gls{ap} scenarios.

\section{Multi-AP Coordination in Wi-Fi 8: Overview and System Model}
\label{sec:mapc}

\begin{figure*}
    \centering
    \includegraphics[scale=0.68]{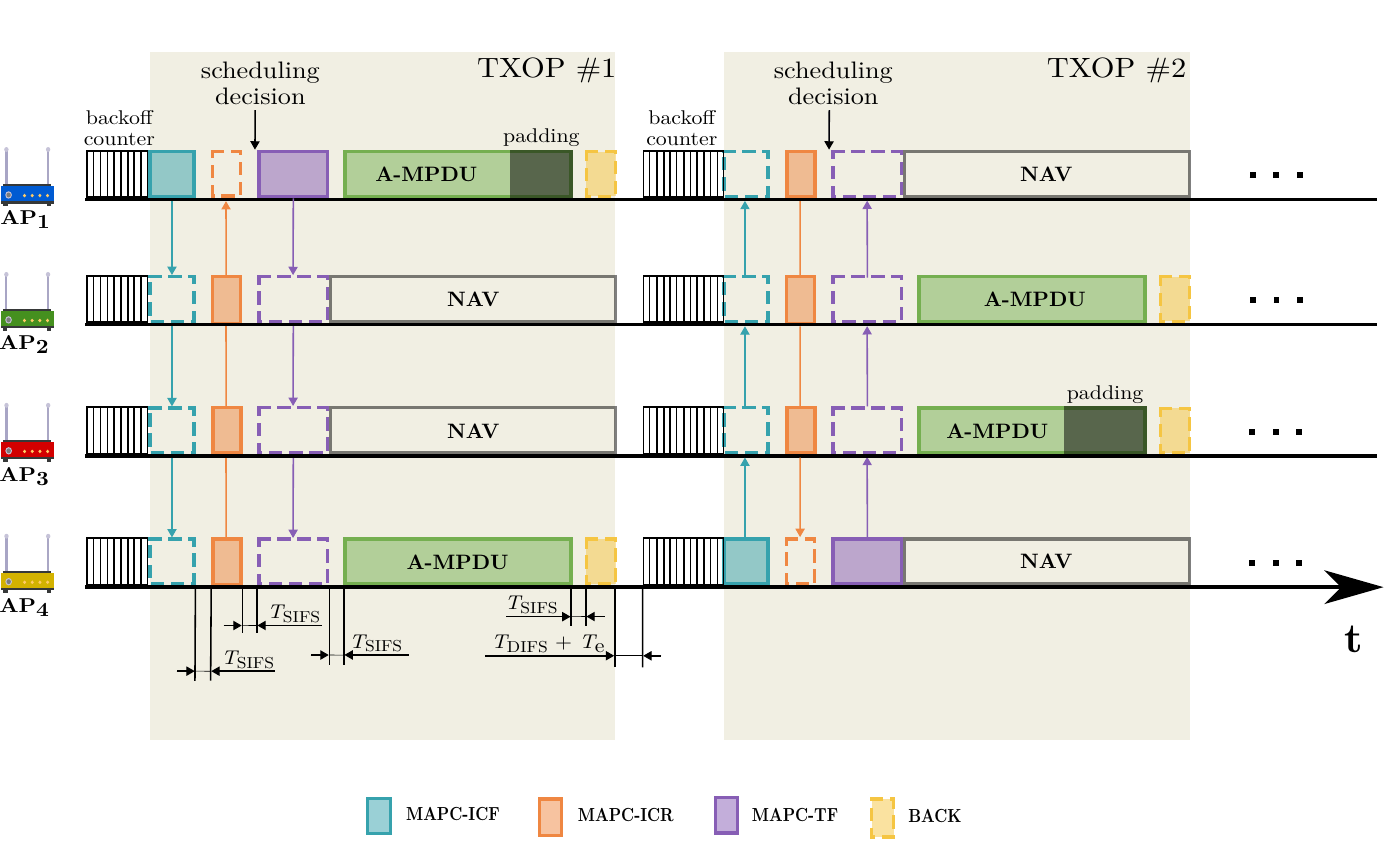}
    \caption{The proposed~\gls{mapc} network.}
    \label{Fig:frames}
\end{figure*}

\subsection{Main Operation}

Based on ongoing discussions about a unified~\gls{mapc} framework for Wi-Fi~8~\cite{tgbn_mapc_vision}, Fig.~\ref{Fig:frames} exemplifies a plausible mechanism for enabling~\gls{csr} in a group of four~\glspl{ap} that can potentially transmit to their corresponding~\glspl{sta}. We consider only downlink transmissions in accordance with the~\gls{mapc} specification in IEEE 802.11bn. The figure illustrates the interaction among the coordinated APs over the course of two \glspl{txop}. All the~\glspl{ap} are within mutual communication range and share a common wireless channel, contending for access using the~\gls{dcf}, which involves selecting a random backoff value and decrementing it while the channel is sensed to be idle before transmitting on the channel. 

To facilitate the coordination among~\glspl{ap}, once the backoff counter reaches zero, the~\gls{ap} that wins the contention acquires the role of Sharing~\gls{ap}---AP$_{1}$ in \gls{txop}~\#1 and AP$_{4}$ in \gls{txop}~\#2 (Fig.~\ref{Fig:frames}). The Sharing~\gls{ap} is responsible for initiating channel reservation by transmitting a~\gls{mapc-icf}. This control frame, among other operations (e.g., indicating the~\gls{mapc} feature to be used), is proposed here to solicit buffer status information from neighboring~\glspl{ap}---called Shared~\glspl{ap}---within the~\gls{mapc} coordination range. In response, these Shared~\glspl{ap} transmit~\gls{mapc-icr} messages, which contain their intention to participate and, for the sake of enabling the mechanism proposed here, also include the number of buffered packets $\varrho_i$ and the timestamp of the \gls{hol} packet $\varepsilon_{i}$ for each of their associated~\glspl{sta}.  \Gls{ofdma} is used to enable the simultaneous transmission of~\gls{mapc-icr} frames. 

While contention-based access enables flexible medium sharing, it can also result in collisions and delay fluctuations. A collision is assumed to occur when no~\gls{mapc-icr} responses are received within a timeout equal to $T_{\rm MAPC\text{-}ICF} + T_{\rm SIFS} + T_{\rm MAPC\text{-}ICR} + T_{\rm DIFS} + T_{\rm e}$, where $T_{\rm SIFS}$ and $T_{\rm DIFS}$ denote the~\gls{sifs} and~\gls{difs}, respectively, and $T_{\rm e}$ represents a backoff slot time. In such cases---indicative of two or more devices in the~\gls{obss} having chosen the same backoff value---each affected device increases its~\gls{cw} and draws a new random backoff value within the updated~\gls{cw} range. The contention process is then restarted using the new backoff value.

On the other hand, if the channel access attempt is successful, the Sharing~\gls{ap} proceeds to parse the received~\gls{mapc-icr} messages and selects the group of~\glspl{sta} to serve in the downlink during the current~\gls{txop}, based on the defined scheduling policy. This decision is made independently at each~\gls{txop}.

Once a group of potential~\gls{sr} transmissions is selected (including multiple~\glspl{ap} and their corresponding~\glspl{sta}), the Sharing~\gls{ap} broadcasts a~\gls{mapc-tf} that identifies the selected~\glspl{sta} (it may contain a single~\gls{sta} if that yields the best outcome according to the scheduler) and specifies their transmission parameters, including bandwidth,~\gls{mcs}, and the duration of the~\gls{txop}.
Following this trigger, the chosen devices transmit their buffered data using an~\gls{ampdu}, while the remaining devices set their \glspl{nav} to defer access until the channel is expected to become idle again. Note that in the proposed framework, channel access follows the legacy contention mechanism, i.e., \gls{dcf}. However, the set of selected devices for transmission does not necessarily include the Sharing~\gls{ap}, as illustrated in \gls{txop} \#2 of Fig.~\ref{Fig:frames}. Finally, each~\gls{sta} sends back its corresponding~\gls{back} upon successful reception. Any frame not successfully received is retained in the buffer and reattempted in the~\gls{sta}'s next scheduled~\gls{txop}.

\subsection{SR Group Creation}\label{subsec:group_creation}

To enable~\gls{csr} in multi-\gls{ap} networks, we precompute a set of feasible~\gls{sr} groups as in~\cite{my_paper_joint_scheduling}.\footnote{Alternative approaches for spatial reuse group creation, such as those presented in \cite{Wojnar_MABS, maks_CoSR_journal}, are also viable. However, we adopt the method proposed in \cite{my_paper_joint_scheduling} to ensure an identical group selection process for both the ML-based models and heuristic algorithms.} Each group consists of a subset of~\gls{ap}–\gls{sta} pairs that can be scheduled for concurrent transmissions, but guaranteeing certain quality in the transmissions. The feasibility of a group is evaluated by estimating the transmission rate used to serve each~\gls{sta} $i$ in the group $n$, which yields
\begin{align}
R^{\rm CoSR}_{i,n} & = \frac{N_{\rm bps} R_{\rm c} N_{\rm sc} N_{\rm ss}}{T_{\rm OFDM} + T_{\rm GI}}\text{,}
\end{align}
where $N_{\rm sc}$, $N_{\rm ss}$, $T_{\rm OFDM}$, and $T_{\rm GI}$ denote the number of data subcarriers, the number of spatial streams used, the duration of an~\gls{ofdm} symbol, and the duration of guard intervals, respectively. The values $N_{\rm bps}$ and $R_{\rm c}$ correspond to the bits per symbol and the coding rate defined by the selected~\gls{mcs} for~\gls{sta} $i$, which is selected on a per-transmission basis according to the expected~\gls{sinr}.

Let $\mathcal{M}_n$ be the subset containing the~\glspl{sta} in group $n$. To assess the feasibility of concurrent transmissions, the performance of each~\gls{sta} $i$ in $\mathcal{M}_n$ is compared to its single transmission case, denoted $R^{\rm ST}_{i}$, which matches with the case in which the~\gls{sta} is served alone (without being exposed to inter-\gls{bss} interference). A group is admitted if the following condition is satisfied:
\begin{align}\label{eq:CSRgroups}
|\mathcal{M}_n| \frac{R^{\rm CoSR}_{i,n}}{R^{\rm ST}_{i}} \geq 1, \quad \forall i \in \mathcal{M}_n \text{.}
\end{align}

Note that the equality in (\ref{eq:CSRgroups}) defines the minimum acceptable transmission rate for an~\gls{sta} to be included in an~\gls{sr} group $n$. This condition ensures that, in the long run, under continuous high traffic loads, the per-\gls{sta} mean throughput is at least as high as that achieved with single transmissions. The expression captures the trade-off between increased interference---due to concurrent transmissions within a group---and the efficiency gained by serving multiple~\glspl{sta} within the same~\gls{txop}. However, satisfying throughput alone does not guarantee that latency requirements are met. Therefore, smart scheduling mechanisms are necessary to effectively manage buffered traffic and minimize delay.

\section{Proposed DRL-based MAPC Scheduling Solution}
\label{sec:ml}

\begin{figure*}
    \centering
    \includegraphics[scale=0.45]{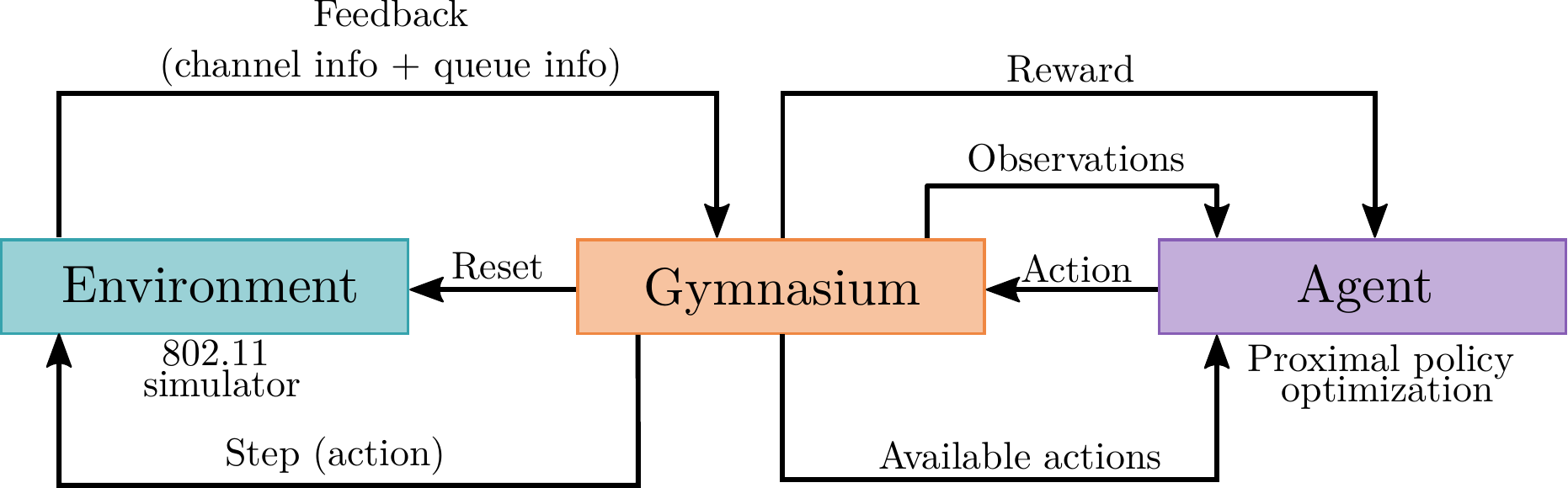}
    \caption{Proposed framework for the integration of~\gls{ml} into 802.11 networks.}
    \label{Fig:implementation}
\end{figure*}

This section provides an overview of the overall solution designed to integrate a~\gls{drl} agent into the Wi-Fi's~\gls{mapc} scheduling operation. As illustrated in Fig.~\ref{Fig:implementation}, we define a system that comprises three core modules: \emph{i)} the network environment, \emph{ii)} the Gymnasium-compatible interface, and \emph{iii)} the learning agent. Next, we define each block in detail.

\subsection{Environment}\label{Sec:environment}

The environment integrates an 802.11 simulator implemented in Python,\footnote{The simulator is available in \url{https://github.com/dncuadrado/11bn-py-Simulator}.} which characterizes particular 802.11 deployments and manages network operations such as traffic arrivals, channel conditions, and transmission events. For the integration of the 802.11 simulator and the~\gls{drl} agent, we consider that, in each episode, a single deployment involving multiple~\glspl{ap} and their associated~\glspl{sta} is simulated for a specific configuration (including a particular random traffic realization). The simulation proceeds in steps, each corresponding to a~\gls{txop}, as illustrated in Fig.~\ref{Fig:frames}, where multiple transmissions from different~\glspl{ap} can be scheduled. Further details on the scenario configuration, channel and traffic models, and scheduling policies are provided in Section~\ref{Sec:evaluation_setup}.

\subsection{Gymnasium}

To enable the training and evaluation of learning-based schedulers, we integrate the 802.11 simulator from Section~\ref{Sec:environment} into the Gymnasium application programming interface (API)---a standardized interface widely used for reinforcement learning environments~\cite{towers2024gymnasium}. This abstraction allows the interaction between the agent and the network environment to follow a consistent loop of observation, action, and reward, while decoupling the learning logic from the simulator's internal implementation.

At the beginning of each episode, the \texttt{reset()} function initializes the environment with a fresh deployment configuration. The initial observation returned to the agent reflects the current network state, including per-\gls{sta} features such as queue size, delay metrics, and channel quality indicators.

As part of the interaction between the Gymnasium and the environment, the \texttt{step()} function is used by the Gymnasium to send the action selected by the agent (Section~\ref{sec:agent}), which corresponds to the index of a valid~\gls{sr} group and is used by the simulator in the upcoming~\gls{txop}. The environment, upon executing the proposed action, updates the queue states and delay metrics, and returns the next observation, the reward signal, and the episode termination flags (\texttt{terminated} and \texttt{truncated}). This feedback enables the agent to learn which scheduling decisions are effective over time.


\subsection{DRL-based MAPC scheduling agent}
\label{sec:agent}

The problem of efficiently scheduling transmissions in an MAPC network can be formulated as a Markov decision process (MDP)~\cite{MDP}, where an agent learns an optimal policy through interactions with the environment. As part of our solution, the agent is trained using the~\gls{ppo} algorithm, implemented in the actor-critic framework~\cite{PPOoriginal_paper}. This approach maintains two distinct neural networks: an actor network, which maps the current observation $s$ to a probability distribution over actions $a$, and a critic network, which estimates the expected return from observation $s$. Both networks are optimized simultaneously to improve policy quality while ensuring stable learning dynamics. The core components of this~\gls{drl} framework are the observation space, the action space, and the reward function, which are elaborated below.

\subsubsection{Observation Space}

At each decision step with time\-stamp $t$, the agent receives an observation vector $s$ containing normalized features from all $N$~\glspl{sta} in the network. The resulting observation vector,
$s = [\boldsymbol{\bar{ \delta}}, \boldsymbol{\bar{\varrho}}, \boldsymbol{\bar{h}} ]  \in [0,1]^{3N}$, is composed of the following information:
\begin{itemize}
    \item \textbf{Delay vector} $\boldsymbol{\bar{ \delta}} = [\delta_1, \dots, \delta_N] / T_{\text{sim}}$, where $\delta_i = (t - \varepsilon_{i})$, $i = 1, ..., N$, is the per-\gls{sta} elapsed time since the~\gls{hol} packet arrived to their corresponding~\gls{ap} queue, normalized by episode duration, $T_{\text{sim}}$. 
    \item \textbf{Queue size vector} $\boldsymbol{\bar{\varrho}} = [\varrho_1, \dots, \varrho_N]/\varrho_{\text{max}}$, which represents the current occupancy of the transmission queue for each~\gls{sta}, $\varrho_{i}$ (in number of frames) normalized by the maximum buffer length, $\varrho_{\text{max}}$.
    \item \textbf{Channel coefficient vector} $\boldsymbol{\bar{h}} = [h_1, \dots, h_N]/h_{\text{max}}$, which includes the per-\gls{sta} link quality, i.e., the channel coefficients normalized by the path loss at a 1~meter distance in~\gls{los}, $h_{\text{max}}$.
\end{itemize}

The considered observations are passed through a shared feature extractor composed of a \gls{mlp} of two layers, where each hidden layer contains \num{64} units and employs the hyperbolic tangent (\texttt{Tanh}) activation function. The resulting feature vector is used as input to two output heads~\cite{actor_critic_original_paper}:

\begin{itemize}
    \item The \textbf{actor} computes action logits, which are transformed into a probability distribution over a discrete action space via a categorical distribution. 
    \item The \textbf{critic} outputs a scalar value that estimates the expected cumulative reward from observation $s$, guiding the policy update by reducing variance in the gradient estimates.
\end{itemize}

\subsubsection{Action Space and Masking}

The discrete action space $\mathcal{A}$ is composed of $Z > 0$ actions, i.e., valid~\gls{sr} groups. For the sake of efficiency, at each step, a binary mask $m \in \{0,1\}^{Z}$ dynamically filters out infeasible or non-beneficial actions.\footnote{Implemented via \texttt{MaskablePPO} from the \texttt{sb3\_contrib} library: \url{https://github.com/Stable-Baselines-Team/stable-baselines3-contrib}.} An action $a_z$ is masked (i.e., $m_z = 0$) if any of the following conditions apply:
\begin{enumerate}
    \item \textit{Offline filtering:} The group is preemptively discarded for the entire deployment realization if it contains at least one receiver that does not meet the condition in \eqref{eq:CSRgroups}.
    \item \textit{Online masking:} At each scheduling decision, the group is masked if none of the intended receivers have pending packets in their respective transmitter queues.
\end{enumerate} 

Masking ensures that the agent selects only from a subset of valid and effective scheduling actions at each step, which considerably reduces the exploration space and accelerates convergence by guiding the agent away from invalid or non-productive actions.

\subsubsection{Reward Design}

To guide the agent toward reducing delays and accelerating learning, we construct a reward function that balances long-term feedback with immediate signals through reward shaping as follows:
\[
r = r_{\text{sh}} + r_{\text{lg}}\text{,}
\]
where $r_{\text{sh}}$ is a reward shaping component that delivers immediate feedback for selecting actions aligned with the current network state. In contrast, $r_{\text{lg}}$ reflects the cumulative effect of previous decisions on the system's overall delay. While $r_{\text{lg}}$ is provided at every step (i.e., it is not sparse), its magnitude diminishes sharply when the worst-case queueing delay increases. Both terms are discussed in the following.

\textbf{Reward shaping:} This component allows the agent to recognize effective actions even when the overall delay remains high. It contributes a positive reward when the head-of-line packet, observed across all~\glspl{sta} at the previous decision point, is successfully transmitted during the current~\gls{txop}, resulting in: 
\[
r_{\text{sh}} = \min \{\boldsymbol{\varepsilon}\} - \min \{\boldsymbol{\varepsilon}'\}\text{,} 
\]
where $\boldsymbol{\varepsilon}' = [\varepsilon'_1, \dots, \varepsilon'_N]$ and $\boldsymbol{\varepsilon} = [\varepsilon_1, \dots, \varepsilon_N]$, denote the head-of-line delays before and after the current~\gls{txop}, respectively. If the same packet remains at the head of the queue, indicating it was not dispatched, then $r_{\text{sh}} = 0$.      

\textbf{Long-term reward:}
This term encourages the agent to maintain queueing delays low in the long run:
\[
r_{\text{lg}} = \min \left( \frac{\beta}{t - \min \{\boldsymbol{\varepsilon}\} + \nu}, 1 \right),
\]
where $\beta > 0$ is a scaling parameter, and $\nu > 0$ prevents indetermination in division by zero. The denominator captures the maximum system-wide queueing delay, with the term decreasing as delays increase.
 
At early stages of training, the knowledge collected by the agent might still be insufficient to determine the best actions and, thereby, delay tends to accumulate. The term $r_{\text{sh}}$ enables the agent to recognize specific actions (such as clearing the oldest buffered packet) leading to meaningful improvements. As the policy matures, $r_{\text{lg}}$ becomes more informative, encouraging sustainable low-delay operation. The combined reward, therefore, guides the learning process from initial exploration toward effective long-term scheduling strategies.

\section{Evaluation Setup}\label{Sec:evaluation_setup}

In this section, we describe the details of our IEEE 802.11bn simulations and the considered scheduling techniques. The simulation parameters are summarized in Table~\ref{tab:simulation_paramters}.

\subsection{Scenario}\label{subsec:scenario}

\begin{figure}
    \centering
    \includegraphics[scale=0.5]{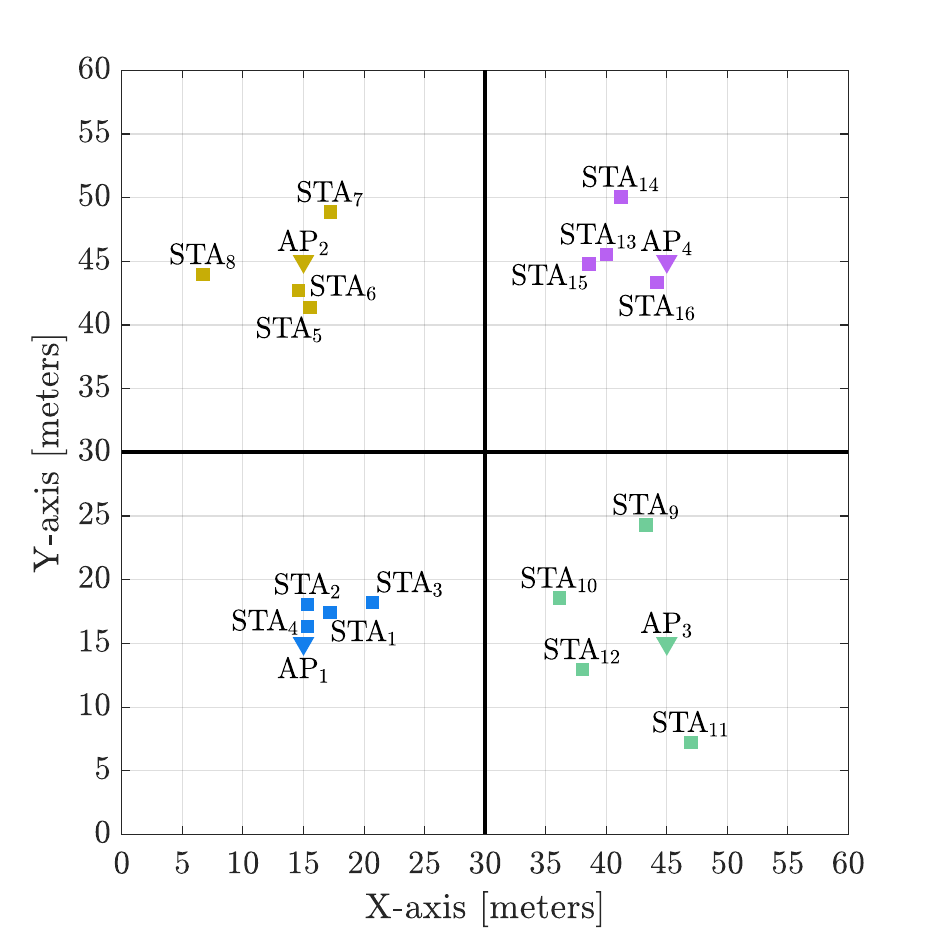}
    \caption{Sample deployment with \num{4}~\glspl{ap} and \num{16}~\glspl{sta}.}
    \label{Fig:scenario}
\end{figure}

We consider an 802.11 enterprise scenario~\cite{pathloss}, where multiple APs coexist within a common area. As illustrated in Fig.~\ref{Fig:scenario}, and for the evaluation of \gls{csr}, we focus only on a subset of $J = 4~$\glspl{ap} operating on the same frequency channel. They are deployed in distinct office rooms separated by walls, with an inter-\gls{ap} distance of \num{30} meters. Each~\gls{ap} $j$ has $N_j$ associated~\glspl{sta}, placed randomly at distances $d_{\text{STA}} \in [1, 10]$ meters from their serving~\gls{ap}. For simplicity, we assume that all~\glspl{ap} have the same number of associated~\glspl{sta}.

Assuming that all the transmissions utilize a fixed transmission power equal to $P_{\text{max}}$, we consider all the~\glspl{ap} to be within the same communication range, so that they contend for the channel according to~\gls{dcf}. Once an~\gls{ap} wins access to the channel, it selects an~\gls{sr} group for transmission. The set of spatially compatible groups available in each deployment is precomputed using the strategy described in Section~\ref{subsec:group_creation}. 

We consider only downlink traffic, with heterogeneous traffic patterns across~\glspl{sta}. Specifically, the traffic profile of each~\gls{sta} is independently drawn from either a Poisson or a bursty traffic model (Section~\ref{subsec:Traffic_Model}), selected with equal probability. Furthermore, the offered load for~\gls{sta} ${i}$, $\omega_i$, is independently and uniformly sampled from the range $[\omega_{\text{min}}, \omega_{\text{max}}]$.
Additionally, only deployments in which at least one scheduling mechanism achieves a 99th-percentile delay below \num{100} ms are included in the results. The remaining cases are considered overloaded and are excluded from the analysis. Such deployments typically arise due to unfavorable \gls{sta} placements or traffic patterns, making them unsuitable for meaningful latency evaluation. The proportion of discarded deployments is reported for each evaluated case.    

\begin{table}
    \caption{Simulation parameters.}
    \label{tab:simulation_paramters}
    \begin{center}
        \begin{tabular}{@{\hspace{0.2cm}}l@{\hspace{0.1cm}}l@{\hspace{0.2cm}}ll}
        \toprule
           \textbf{Parameter}  & \textbf{Description} & \textbf{Value} \\ \midrule
            $d_{\text{STA}}$ & Distance between~\gls{ap} and~\glspl{sta} [meters] & [1, 10] \\
            $B_{\text{p}}$ & Break-point distance [meters]  & 10 \\
            $B$ & Bandwidth [MHz] & 80 \\
            $f_{\text{c}}$ & Carrier frequency [GHz] & 6 \\
            $\sigma$ & Standard deviation of shadowing [dB] & 5 \\
            $N_{\text{SC}}$ & Number of data subcarriers & 980 \\
            $N_{\text{SS}}$ & Number of spatial streams & 2 \\
            $T_{\text{OFDM}}$ & OFDM symbol duration [$\mu$s] & 12.8 \\
            $T_{\text{GI}}$ & Guard interval duration [$\mu$s] & 0.8 \\
            $T_{\text{max}}$ & Max~\gls{txop} duration [ms] & 5 \\ 
            $T_{\text{\gls{mapc}-ICF}}$ &\gls{mapc} Initial Control Frame duration [$\mu$s] & 74.4\\
            $T_{\text{\gls{mapc}-ICR}}$ &\gls{mapc} Initial Control Response duration [$\mu$s] & 88\\
            $T_{\text{\gls{mapc}-TF}}$ &\gls{mapc} Trigger Frame duration [$\mu$s] & 74.4\\
            $T_{\text{BACK}}$ & Block ACK [$\mu$s] & 100 \\
            $T_{\text{SIFS}}$ & Duration of a SIFS slot [$\mu$s] & 16 \\
            $T_{\text{DIFS}}$ & Duration of an DIFS slot [$\mu$s]: & 34 \\
            $T_{\text{e}}$ & Duration of an empty slot [$\mu$s] & 9 \\
            CW$_{\text{min}}$  & Min contention window & 15 \\
            CW$_{\text{max}}$  & Max contention window & 1023 \\
            $P_{\text{max}}$ & Max transmission power [mW] & 200 \\
            $\varrho_{\text{max}}$ & Max buffer size & 10$^{4}$ \\
            $h_{\text{max}}$ & Max channel gain & 10$^{-3}$ \\
            $W$ & Noise power  [Watts] & $3.2 \times 10^{\text{-13}}$  \\
            CCA & Clear channel assessment threshold [dBm] & -82 \\
            $T_{\text{ON}}$ & Bursty traffic ON period mean duration [ms] & 1 \\
            $T_{\text{OFF}}$ & Bursty traffic OFF period mean duration [ms] & 10 \\
            $T_{\text{sim}}$ & Simulation duration [s] & 5 \\
            $L$ & Length of single data frame [bits] & $12\times 10^{\text{3}}$ \\ \bottomrule
        \end{tabular}
    \end{center}
\end{table}

\subsection{Channel Model}

The~\gls{sinr} perceived at~\gls{sta}~$i$, denoted as $\xi_i$, depends on the subset $\mathcal{K}$ of simultaneously active interfering transmitters (if any), and is computed as:
\begin{align}\label{eq:sinr}
\xi_i = \frac{P_j h_{i,j}}{W + \sum_{k \in \mathcal{K}} P_k h_{i,k}} \text{,}
\end{align}
where $P_j$ is the power used by~\gls{ap}~$j$ to deliver the intended transmission to~\gls{sta}~$i$, and $W$ stands for the noise power. The terms $P_k$ represent the power transmitted by interfering~\gls{ap}~$k$, which contributes to the interference experienced at~\gls{sta}~$i$. The channel gain $h_{i,j}$ between~\gls{sta}~$i$ and~\gls{ap}~$j$ is defined as $h_{i,j} = 10^{-\frac{P_{\rm L}(d_{i,j})}{10}}$, a function of the transmitter-receiver separation $d_{i,j}$. Likewise, $h_{i,k}$ represents the channel gain from~\gls{ap}~$k$ to~\gls{sta}~$i$, based on distance $d_{i,k}$. The term $P_{\rm L}(d_{i,j})$ represents the path loss in decibels and is modeled using the TGax specification for enterprise environments~\cite{pathloss}:
\begin{align}\label{eq:pathloss}
P_{\rm L}(d_{i,j}) = & 40.05 + 20\log_{10}\left(\frac{\min(d_{i,j}, B_{\rm p}) f_{\rm c}}{2.4}\right) \nonumber \\ 
& + \mathbbm{1}_{d_{i,j} > B_{\rm p}} P' + 7W_{\rm n} + \mathcal{X} \text{,}
\end{align}
where $d_{i,j} \geq 1$ denotes the distance in meters, $f_{\rm c}$ is the carrier frequency in GHz, $W_{\rm n}$ is the number of intervening walls, and $P' = 35\log_{10}(d_{i,j}/B_{\rm p})$ accounts for extra loss beyond the break-point distance $B_{\rm p}$. The term $\mathcal{X}$ models log-normal shadowing effects with $\ln(\mathcal{X}) \sim \mathcal{N}(0, \sigma)$.    

We adopt the IEEE 802.11be~\gls{mcs} set (no major changes are expected in 802.11bn), and determine the appropriate~\gls{mcs} index for each transmission using pre-generated~\gls{per} against~\gls{snr} curves gathered from MATLAB simulations~\cite{MatlabMCSPERvsSNR}. These mappings reflect the simulation conditions used later in our study, including fixed packet size $L$, bandwidth $B$, channel model, and antenna settings. The selection criterion ensures that only~\gls{mcs} values satisfying ${\rm PER} < 10^{-2}$ for the estimated $\xi_i$ are employed. As a result,~\glspl{sta} closer to their~\glspl{ap} are assigned higher~\gls{mcs} values, while distant~\glspl{sta} operate with more robust (lower)~\gls{mcs} indexes. 

The number of aggregated frames (\gls{ampdu} size) sent to~\gls{sta} $i$ per~\gls{txop} is denoted by $U_i$, which depends on the selected~\gls{mcs} and the~\gls{txop} duration. Interference from concurrent transmissions---i.e., $\sum_{k \in \mathcal{K}} P_k h_{i,k}$ in~\eqref{eq:sinr}---directly affects~\gls{mcs} selection and must be precisely accounted for when grouping~\glspl{sta} for~\gls{sr} compatibility.

The number of~\glspl{mpdu} successfully decoded by~\gls{sta} $i$ is modeled as a binomial random variable, i.e., $\mu_i \sim B(U_i, q)$, where $q = 1 - {\rm PER}$ is the success probability and $U_{i}$ the number of transmitted frames. Frames that~\gls{sta} $i$ fails to decode are retained in its corresponding~\gls{ap}'s buffer and are re-attempted\footnote{We assume that retransmissions are not subject to any maximum attempt limit.} during the next~\gls{txop} scheduled for~\gls{sta}~$i$.


\subsection{Traffic Model}\label{subsec:Traffic_Model}

Downlink traffic is considered, and each~\gls{ap}$_j$ maintains independent logical transmission queues for its associated $N_j$~\glspl{sta}, with incoming packets being buffered. We assume that each~\gls{sta} supports a single flow, and $\varrho_i(t)$ denotes the queue depth for~\gls{sta} $i$ at time $t$. For each~\gls{txop}, the queue dynamics evolve according to
\begin{align}\label{eq:queuesize}
\varrho_i(t + T_{\rm s}) = \max\left\{\varrho_i(t) - \mu_i(T_{\rm s}), 0\right\} + \lambda_i(T_{\rm s}) \text{,}
\end{align}
where $\mu_i(T_{\rm s})$ is the number of successfully delivered frames during a~\gls{txop} of duration $T_{\rm s}$ (drawn from a random variable with expected value $\bar\mu$), 
and $\lambda_i(T_{\rm s})$ represents the number of arrivals during that same interval (drawn from a random variable with expected value $\bar\lambda$).

The following two traffic generation models are used to evaluate system performance, and in all cases, the load at~\gls{sta}~$i$ is defined as $\omega_i$ [Mb/s].

\begin{enumerate}
    \item \textit{Poisson}: Packet arrivals follow a Poisson process, resulting in exponentially distributed inter-arrival times for each~\gls{sta}.
    \item \textit{Bursty}: This model captures intermittent traffic where packet bursts occur in ON periods separated from OFF periods. Burst arrivals are modeled as a two-state (ON/OFF) Markovian process, with ON and OFF durations drawn from exponential distributions with means $T_{\rm ON}$ and $T_{\rm OFF}$, respectively. During the ON periods, the traffic load aggregates all the arrivals from an equivalent Poisson process.
\end{enumerate}   

We assume equal priority for all packets within each~\gls{sta} queue and employ a~\gls{fifo} scheduling policy for managing each queue.


\subsection{MAPC Scheduling Mechanisms}

We evaluate the performance of five different~\gls{mapc} scheduling mechanisms, comprising three heuristics and two learning-based agents:

\begin{itemize}
    \item \textit{Maximum Number of Packets (\gls{mnp}):}
    This scheduler selects the~\gls{sr} group with the largest number of packets in the current~\gls{txop}. For each feasible group, the number of schedulable packets is estimated based on the current queue lengths, regardless of packet age.
    \item \textit{Oldest Packet (\gls{op}):}
    The~\gls{op} policy prioritizes the~\gls{sr} group that includes the~\gls{sta} with the highest~\gls{hol} packet delay. The goal is to reduce the worst-case delay by targeting the most delayed flow at each decision step. This mechanism implicitly promotes fairness but may underutilize~\gls{txop} capacity when the most delayed flows cannot be scheduled concurrently.
    \item \textit{Traffic-Alignment Tracker (\gls{tat}):}
    Initially proposed in~\cite{my_paper_joint_scheduling},~\gls{tat} seeks to balance worst-case delay reduction and efficiency by selecting the~\gls{sr} group that minimizes a delay-alignment cost, i.e., the difference between the arrival times of the \gls{hol} packets among all the participants in a group. As a result,~\gls{tat} tends to maintain stable delay distributions across heterogeneous deployments.
    \item \textit{Machine Learning-General Agent (ML-G):}
    This~\gls{drl}-based scheduler is trained to generalize across a wide range of traffic and deployment scenarios. At each decision step, the agent observes normalized queue lengths,~\gls{hol} delays, and channel coefficients for all~\glspl{sta}, and selects a valid~\gls{sr} group using a policy trained via the~\gls{ppo} algorithm~\cite{PPOoriginal_paper}. The agent’s objective is to minimize the system-wide worst-case delay across all packets and~\glspl{sta}. Action masking is used to dynamically filter out infeasible or non-beneficial~\gls{sr} groups.
    \item \textit{Machine Learning-Expert Agent (ML-E):}
    ML-E shares the same architecture, observation space, and masking procedure as ML-G, but it is trained under a specialized setting: a fixed deployment topology (corresponding to Fig.~\ref{Fig:scenario}) with diverse traffic realizations. This focused training enables the agent to fine-tune its scheduling strategy to the spatial structure and interference patterns of a single environment, while still generalizing over different traffic conditions. 
\end{itemize}


\section{Performance Evaluation}
\label{sec:performance}

This section delves into the performance evaluation of the proposed~\gls{drl}-based scheduling mechanism. We first show the process of training the~\gls{ml} schedulers and then examine their performance in new (unseen) deployments.

\subsection{Training}

The training of the~\gls{drl} agent is conducted using the~\gls{ppo} algorithm, implemented via the \texttt{MaskablePPO} from the \texttt{sb3\_contrib} library~\cite{stable_baselines3}. The agent is trained within the custom Gymnasium environment described earlier, where each training episode includes the simulation of a Wi-Fi deployment, including~\gls{sta} placements with their corresponding channel conditions and the generation of new traffic patterns with per-STA load of $\omega_{i}$ [Mb/s], which remain fixed throughout the episode. Given that the channel model accounts solely for path loss (as a function of transmitter-receiver distance) and shadowing, we assume the channel conditions remain static throughout each episode. Moreover, the duration of each episode is the same as the duration of the simulation, that is, $T_{\text{sim}}$.

During the training of the model, we apply cosine annealing to gradually decay the learning rate over time, which is helpful to improve policy generalization and performance. The model is periodically evaluated, and training is halted when the agent converges to a stable policy (i.e., no improvements are observed after 20 successive evaluations) or upon reaching a predefined number of environment steps, $n_{\text{total}}$. Model checkpoints and performance metrics are logged using the Weights \& Biases platform.\footnote{\url{https://wandb.ai/site}} 
For better computational efficiency and faster convergence, we employ parallelized training using the \texttt{SubprocVecEnv} wrapper to instantiate multiple independent environments~\cite{stable_baselines3}. According to this, the agent interacts with a total of $n_{\text{env}}$ parallel environments during each rollout phase. The values of the main hyperparameters used in training after tuning are reported in Table~\ref{tab:training_parameters}. 

\begin{figure}
    \centering
    \includegraphics[scale=0.47]{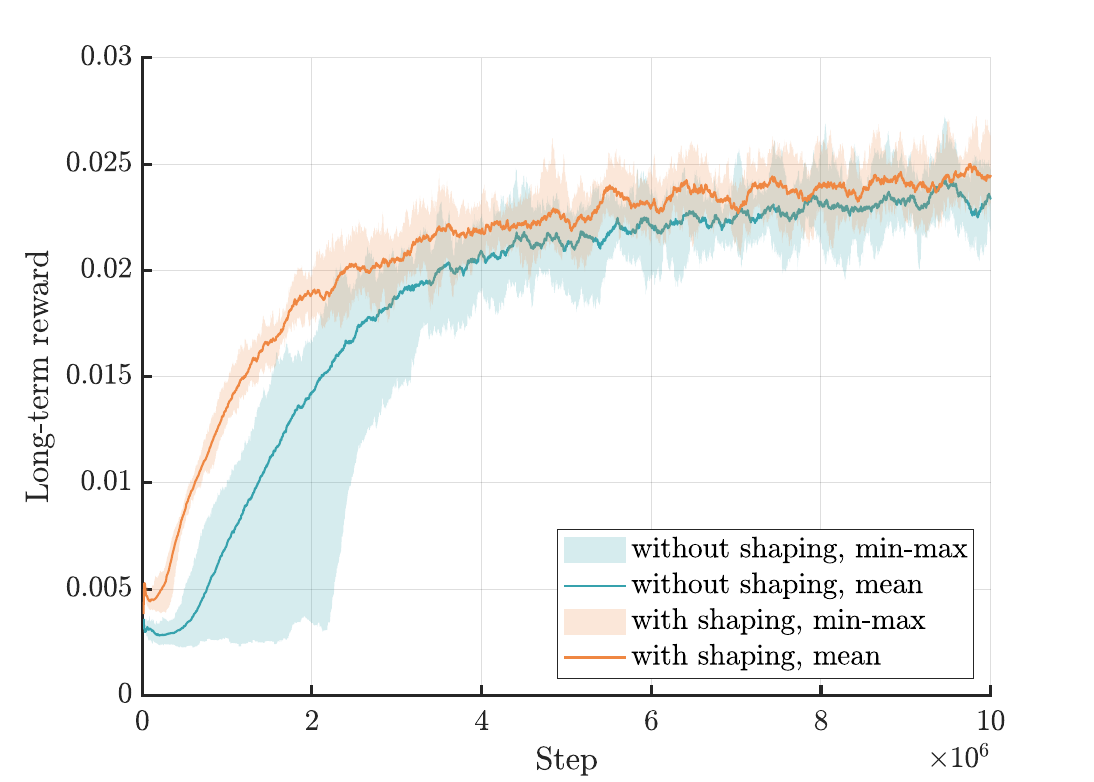}
    \caption{Long-term reward evolution during an ML-G agent training with/without reward shaping.}   \label{Fig:long_term_reward_evolution}
\end{figure}

Next, to assess the importance of the reward function, Fig.~\ref{Fig:long_term_reward_evolution} shows the long-term reward evolution of two ML-G agents---one trained with reward shaping and one without---each evaluated across 10 independent training runs of $10^7$ steps. As shown, the agent using reward shaping exhibits significantly faster initial learning, achieving higher rewards within the first 4 million steps. This indicates that the additional feedback guides early exploration toward more promising scheduling actions. While both agents eventually converge to similar performance levels around 6–10 million steps, the shaped reward continues to provide a slight advantage in terms of final training performance stability and peak reward. These results suggest that reward shaping not only accelerates convergence but also leads to better policy refinement in later training stages.

\begin{figure}
    \centering
    \includegraphics[scale=0.47]{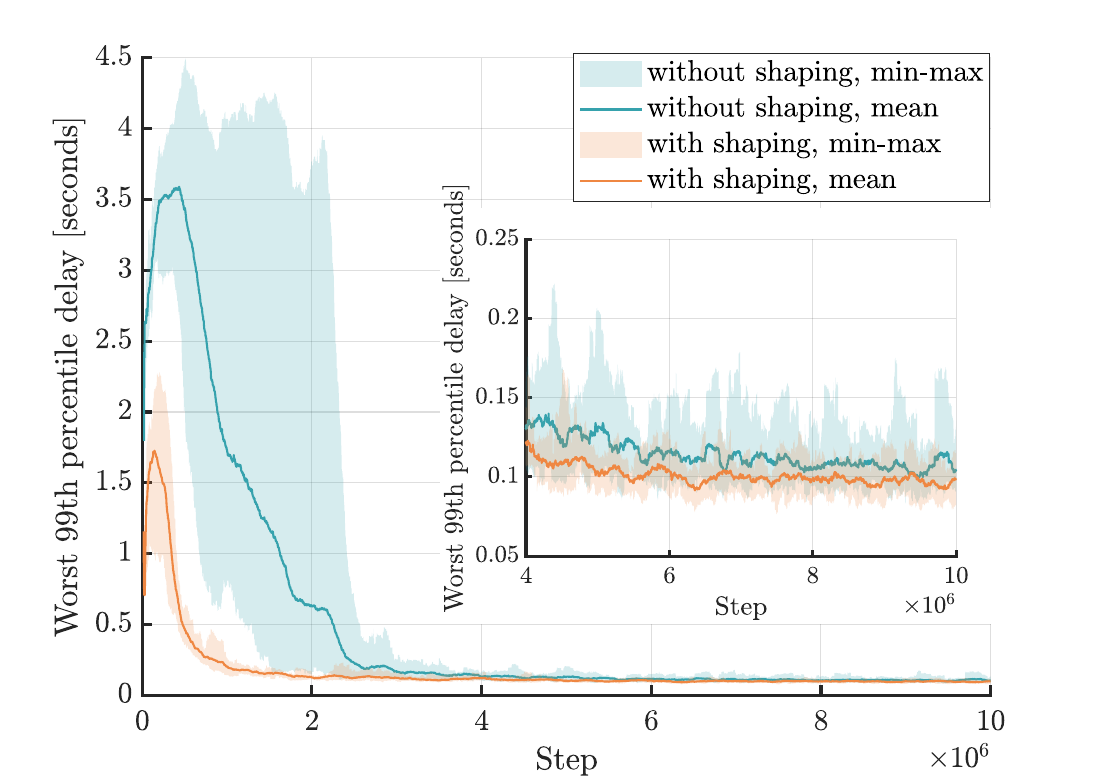}
    \caption{Evolution of the worst $99^{\text{th}}$-percentile delay during an ML-G agent training with/without reward shaping.}   
    \label{Fig:worst99percentile_evolution}
\end{figure}

Analogously, Fig.~\ref{Fig:worst99percentile_evolution} shows the actual worst-case delay performance (as the worst $99^{\text{th}}$-percentile delay of each episode across all~\glspl{sta}) of the two agents when trained with and without reward shaping. The shaded regions represent the min–max bounds over the same 20 trainings previously analyzed (10 with and 10 without reward shaping), while the solid curves show the mean values. Both agents progressively reduce delay as training advances; however, the model with reward shaping converges significantly faster and reaches lower worst-case delay values throughout the entire training process. The agent without shaping exhibits higher variability and slower convergence, particularly during early stages. The inset plot on the bottom right provides a close-up of the stable region (from $4\times 10^6$ to $10^7$ steps), where the advantage of reward shaping remains consistent. 

\begin{table}
    \caption{Training parameters.}
    \label{tab:training_parameters}
    \begin{center}
        \begin{tabular}{@{\hspace{0.15cm}}l@{\hspace{0.2cm}}l@{\hspace{0.2cm}}ll}
        \toprule
           \textbf{Parameter}  & \textbf{Description} & \textbf{Value} \\ \midrule
            $\texttt{gamma}$ & Discount factor & 0.99 \\
            $\texttt{gae\_lambda}$ & GAE (Generalized Advantage Estimation) & 0.92 \\
            $\texttt{n\_steps}$ & Steps per environment per update & 128 \\
            $\texttt{batch\_size}$ & Minibatch size & 256 \\
            $\texttt{clip\_range}$ & Clipping parameter & 0.2 \\
            $\alpha_{\text{init}}$ & Initial learning rate & $6.5 \times 10^{-4}$ \\
            $\alpha_{\text{schedule}}$ & Learning rate schedule & Cosine decay \\
            $n_{\text{env}}$ & Number of parallel environments & 10 \\ 
            $n_{\text{total}}$ & Total number of steps & $ 10^7$ \\   
            $\beta$ & Scale factor for long-term reward & $10^{-3}$ \\ 
            $\nu$ & Factor to avoid indetermination & $10^{-6}$ \\ \bottomrule
        \end{tabular}
    \end{center}
\end{table}

\subsection{Inference Results}

To validate our proposal, we first evaluate the performance of our trained~\gls{drl} agents, as well as the heuristic algorithms, in the deployment shown in Fig.~\ref{Fig:scenario}. Then, we extend the experiment to other randomly generated deployments to verify the adaptability and generalization of our proposal. Finally, we analyze how the ML-based agents scale as the number of users increases from $N = 8$ to $N = 20$.

\subsubsection{Sample Deployment}

\begin{figure}
    \centering
    \includegraphics[scale=0.47]{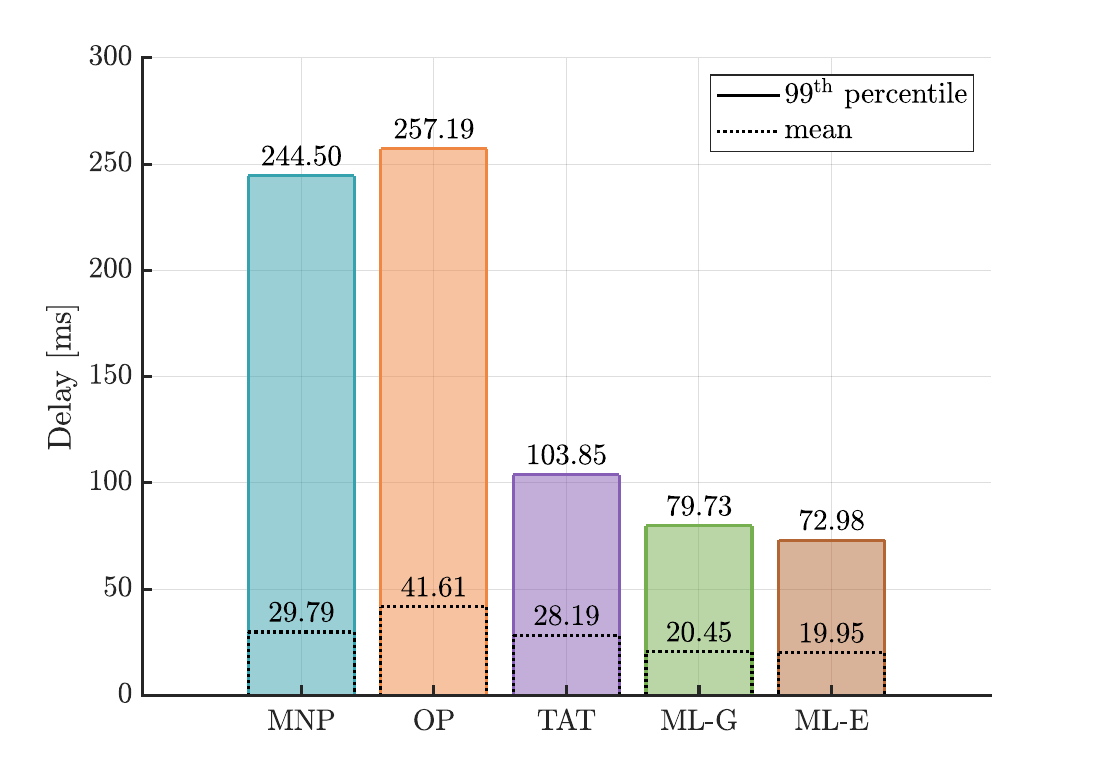}
    \caption{$99^{\text{th}}$-percentile and mean delay for all the scheduling strategies in the sample deployment, after 100 traffic realizations, with $\omega_{i} \in [10,90]$ and 10\% of traffic realizations discarded.}
    \label{Fig:sample_deplo_10_90}
\end{figure}

Fig.~\ref{Fig:sample_deplo_10_90} presents the $99^{\text{th}}$-percentile and mean delays achieved by all the evaluated schedulers for 100 different traffic realizations in the deployment shown in Fig.~\ref{Fig:scenario}, with \num{4}~\glspl{ap} and \num{16}~\glspl{sta}. Every~\gls{sta} is independently assigned either a Bursty or Poisson traffic flow in each traffic realization, and its offered load $\omega_i$ is randomly drawn from the range $[10, 90]$---which corresponds to a mean load of \num{50} Mb/s, comparable to the medium-high traffic regime discussed later in this section. Among the heuristic baselines,~\gls{mnp} and~\gls{op} exhibit the highest worst-case delay, exceeding 240 ms. This confirms their poor adaptability in highly loaded or Bursty scenarios, where prioritizing either the aggregate queue length (\gls{mnp}) or the head-of-line packet (\gls{op}) can lead to starvation and inefficient group selections. In contrast,~\gls{tat}, which explicitly balances traffic alignment and efficient transmissions, significantly reduces the $99^{\text{th}}$-percentile delay to 103.85 ms.

Both learning-based schedulers, ML-G and ML-E, outperform all heuristics by a wide margin. Notably, ML-E achieves the lowest worst-case delay of 72.98 ms, representing a reduction of 8\% compared to ML-G, a 30\% improvement over~\gls{tat}, and over 70\% compared to~\gls{mnp} and~\gls{op}. These results suggest that the~\gls{drl} agents effectively schedule groups under diverse traffic variations while balancing the worst-case delay and efficient transmissions. In both scenarios, the~\gls{ml}-based schedulers consistently achieve lower mean delay compared to all heuristic-based mechanisms. 

\begin{figure*}
    \centering
    \includegraphics[scale=0.55]{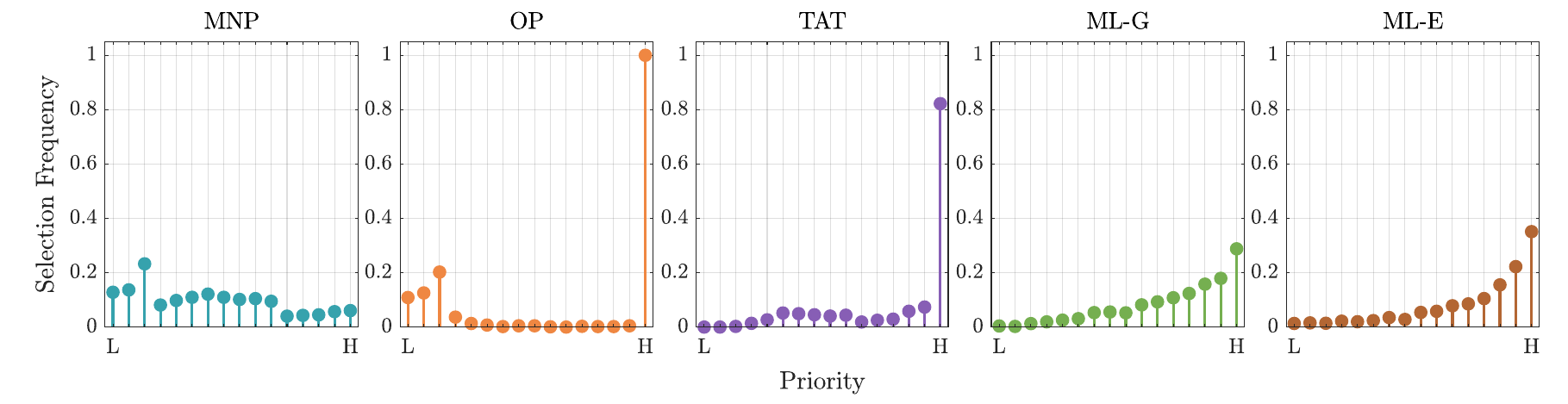}
    \caption{Normalized selection frequency per priority index for each evaluated scheduling strategy in the deployment of Fig.~\ref{Fig:scenario} and one traffic realization.}
    \label{Fig:selection_frequency}
\end{figure*}

Fig.~\ref{Fig:selection_frequency} presents the normalized selection frequency per priority index for each evaluated scheduling strategy, measured in the deployment depicted in Fig.~\ref{Fig:scenario} with one traffic realization. The priority index ranges from low (L) to high (H), where L corresponds to selected actions that contain---at least---the~\gls{sta} with the lowest~\gls{hol} delay, and H to those with the highest. Note that since buffer states continuously evolve as transmissions occur, the same action may correspond to different priority levels across successive~\glspl{txop}. The y-axis shows the fraction of~\glspl{txop} in which each priority level was selected. Note that~\gls{mnp} distributes its selections relatively evenly, reflecting its throughput-oriented design, while~\gls{op} always concentrates on the highest priority, consistently serving the oldest packets, but at the cost of serving also~\glspl{sta} with low priority that also belong to the selected~\gls{sr} group.~\gls{tat} also favors high priorities, though with a smoother distribution that reflects its delay-balancing mechanism. In contrast, ML-G exhibits a gradual increase across the priority spectrum, indicating a learned policy that balances delay reduction with broader scheduling diversity. Similarly, ML-E shows a high diversity but slightly sharper bias toward high-priority levels, due to its specialization in this deployment.

\subsubsection{Random Deployments}

\begin{figure*}
        \centering
        \begin{subfigure}{0.32\textwidth}
            \centering
            \includegraphics[scale=0.36]{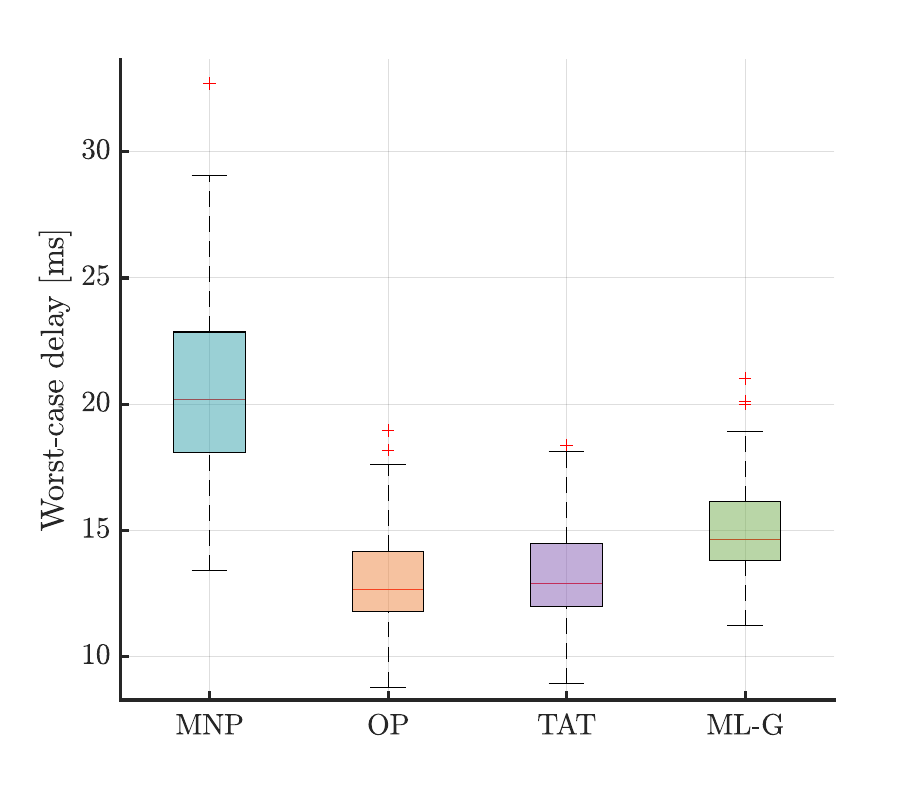}
            \caption{Low load}
            \label{fig:random_deplo_boxplot_10_30}
        \end{subfigure}
        \hfill
        \begin{subfigure}{0.32\textwidth}
            \centering
            \includegraphics[scale=0.36]{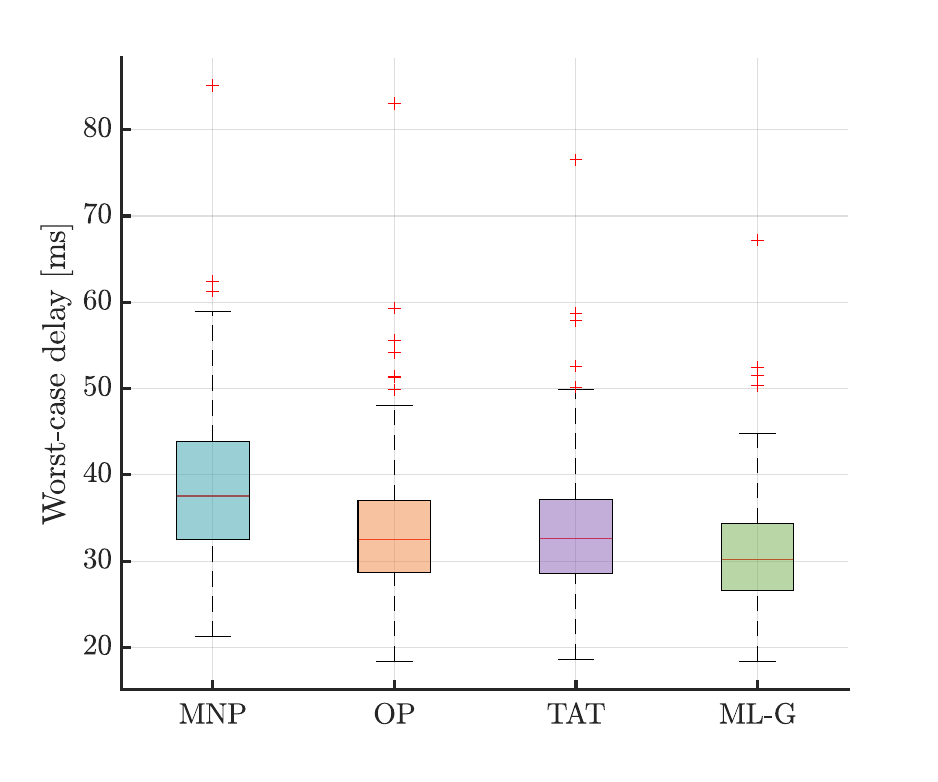}
            \caption{Medium load}
            \label{fig:random_deplo_boxplot_30_50}
        \end{subfigure}
        \hfill
        \begin{subfigure}{0.32\textwidth}
            \centering
            \includegraphics[scale=0.36]{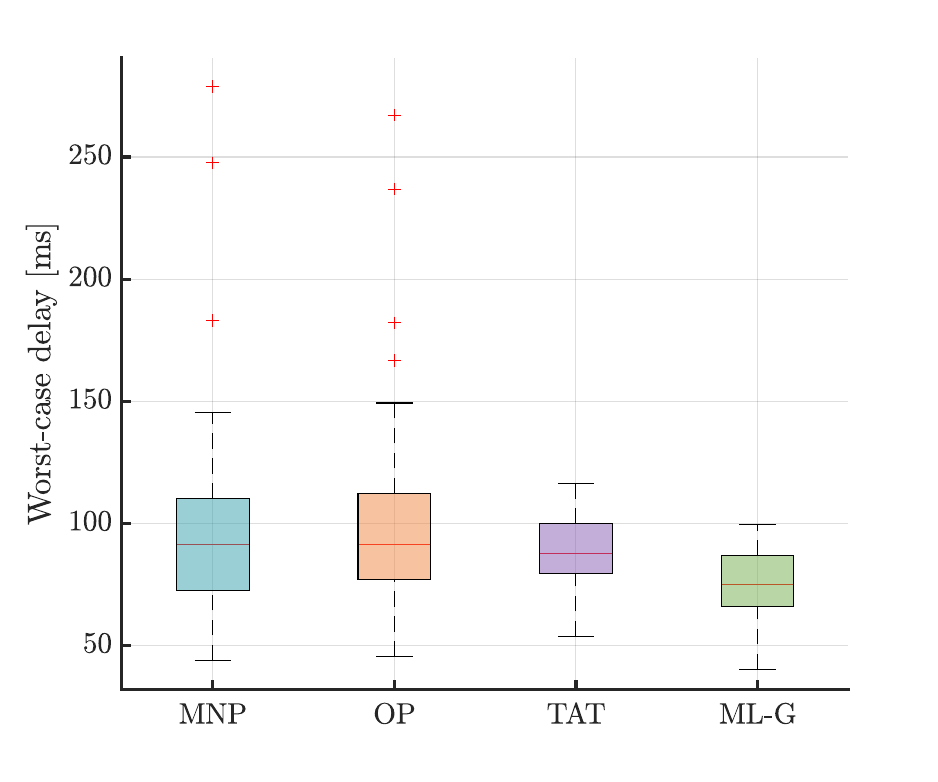}
            \caption{High load}
            \label{fig:random_deplo_boxplot_50_70}
        \end{subfigure}
        \caption{Worst-case delay distribution over 100 random deployments.}
        \label{fig:random_deplo_boxplot}
    \end{figure*}

Fig.~\ref{fig:random_deplo_boxplot} provides a detailed performance analysis by showing the distribution of worst-case delays across 100 random deployments. In each deployment, \num{4}~\glspl{ap} are positioned as in Fig.~\ref{Fig:scenario}, while $N_j = 4$~\glspl{sta} per~\gls{ap} (totaling $N = 16$ stations) are randomly placed around their corresponding~\glspl{ap}. We evaluate all scheduling strategies under three per-\gls{sta} traffic load regimes, low: $\omega_i \in~[10,30]$, medium: $\omega_i \in [30,50]$, and high: $\omega_i \in [50,70]$. For low and medium traffic loads, all deployments are included in the analysis. Under high load conditions, however, 41\% of the deployments are excluded due to network overload. Results for ML-E are omitted in this setting, as its performance degrades significantly when evaluated on deployments different from the one used during training. 

Under low load, Fig.~\ref{fig:random_deplo_boxplot_10_30}, all schedulers maintain bounded delay, but~\gls{mnp} displays significantly higher variability, occasionally producing worst-case delays over 30 ms. The~\gls{ml}-based method exhibits tighter delay bounds overall, but does not outperform~\gls{op} or~\gls{tat} under low-load conditions, with~\gls{op} emerging as the most effective scheduler in this regime, because the network is underutilized and the optimal solution is to serve the HoL packet.

As the load increases, Fig.~\ref{fig:random_deplo_boxplot_30_50}, non-ML schedulers begin to diverge.~\gls{mnp} and~\gls{op} show higher median delay and a broader spread of outliers.~\gls{tat} remains more stable, but ML-G consistently outperforms all baselines, with a narrower interquartile range and fewer outliers, because it balances addressing the worst-case differently, not always dispatching the oldest packet, as shown in Fig.~\ref{Fig:selection_frequency}. 

Under high load, Fig.~\ref{fig:random_deplo_boxplot_50_70}, the performance gap between ML-G and baselines widens substantially.~\gls{mnp} and~\gls{op} experience frequent worst-case delays above 100 ms.~\gls{tat} degrades more gracefully but still trails the~\gls{ml} agent. ML-G maintains both a lower median and a tighter delay distribution, as it adopts a more flexible strategy for handling delay, rather than strictly serving the oldest packet at every decision step, confirming the learned policy’s resilience under network stress.

\begin{figure}
    \centering
    \includegraphics[scale=0.47]{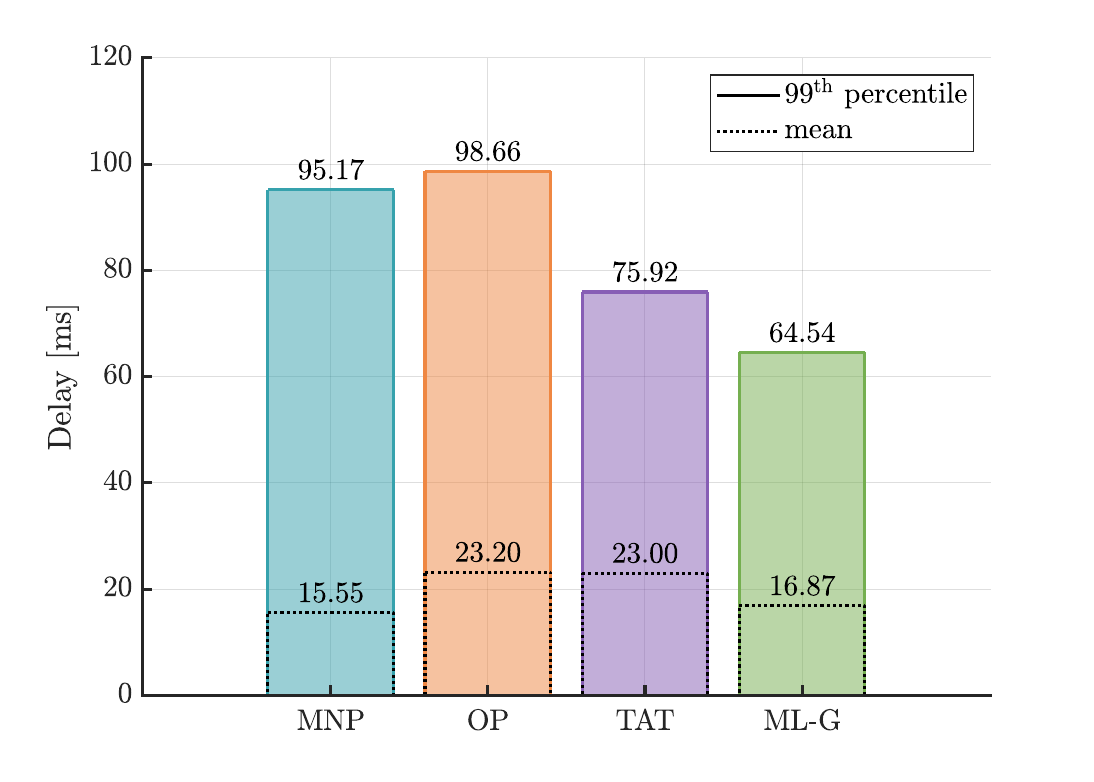}
    \caption{$99^{\text{th}}$-percentile and mean delay for all the scheduling strategies after 100 deployments realizations, with $\omega_{i} \in~ [10,90]$ and 15\% of deployments discarded.}
    \label{Fig:random_deplo_10_90}
\end{figure}

Finally, Fig.~\ref{Fig:random_deplo_10_90} presents the 99th-percentile and mean delay achieved by the four scheduling strategies---\gls{mnp},~\gls{op},~\gls{tat}, and ML-G---across 100 deployment realizations, where $\omega_i \in [10, 90]$---the same range used during training of the ML-G agent---and 15\% of the deployments are excluded from the analysis due to overload. The results demonstrate that the proposed ML-based scheduler (ML-G) outperforms all heuristic baselines in terms of both mean and worst-case delay. Specifically, ML-G achieves the lowest 99th-percentile delay at 64.54 ms, representing a 15\% improvement over~\gls{tat} (75.92 ms), the best-performing heuristic policy. In terms of mean delay, ML-G also performs competitively, reaching 16.87 ms---slightly higher than~\gls{mnp}’s 15.55 ms but significantly lower than~\gls{op} (23.20 ms) and~\gls{tat} (23.00 ms).

These results collectively demonstrate that the learning-based scheduler not only outperforms heuristic baselines in average and tail delay metrics but also exhibits superior stability across different load conditions. Its ability to capture cross-scenario scheduling patterns gives it a distinct advantage over fixed-priority policies.

\subsubsection{Scalability}

\begin{figure*}[t]
    \centering

    \begin{subfigure}{0.24\textwidth}
        \centering
        \includegraphics[width=\linewidth]{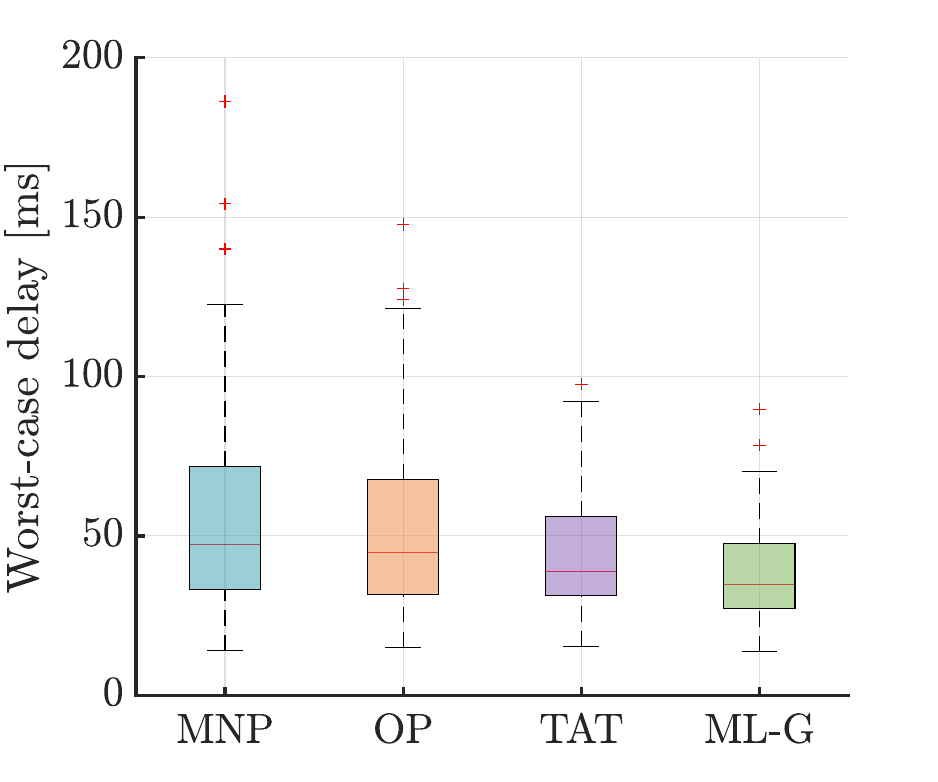}
        \caption{$N = 8$~\glspl{sta}}
        \label{fig:scalability_8stas}
    \end{subfigure}
    \hfill
    \begin{subfigure}{0.24\textwidth}
        \centering
        \includegraphics[width=\linewidth]{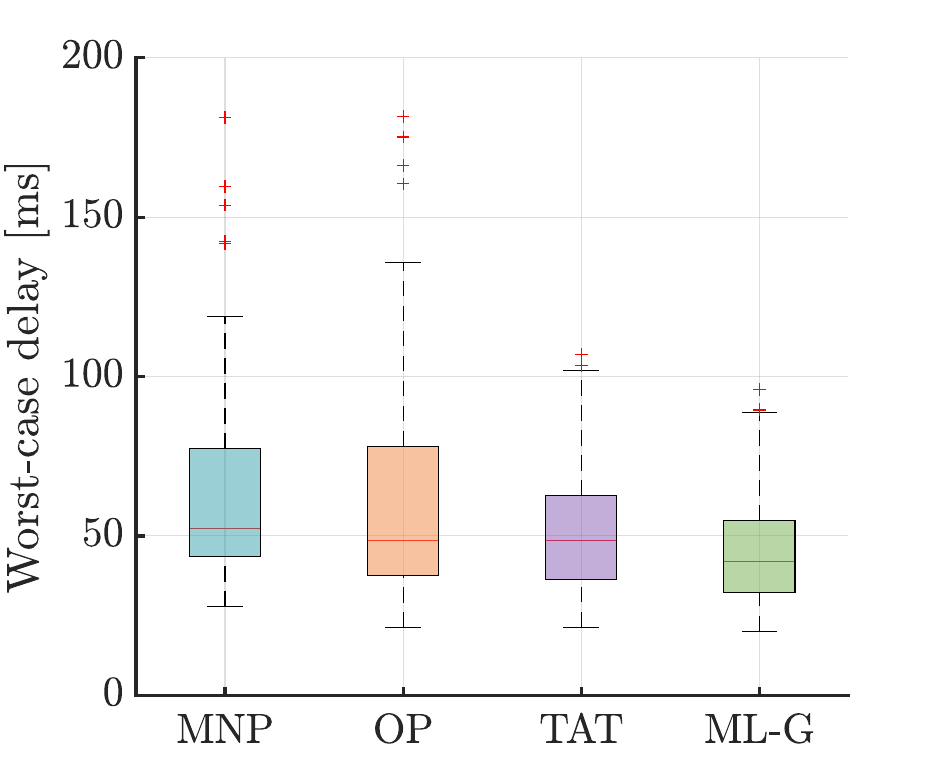}
        \caption{$N = 12$~\glspl{sta}}
        \label{fig:scalability_12stas}
    \end{subfigure}
    \hfill
    \begin{subfigure}{0.24\textwidth}
        \centering
        \includegraphics[width=\linewidth]{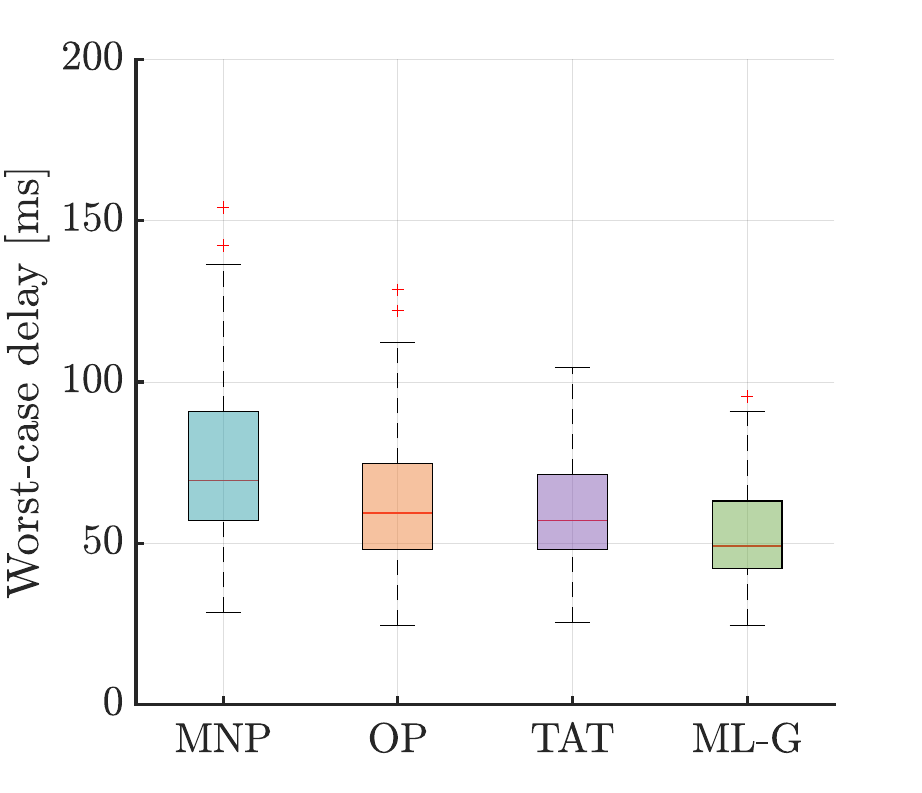}
        \caption{$N = 16$~\glspl{sta}}
        \label{fig:scalability_16stas}
    \end{subfigure}
    \hfill
    \begin{subfigure}{0.24\textwidth}
        \centering
        \includegraphics[width=\linewidth]{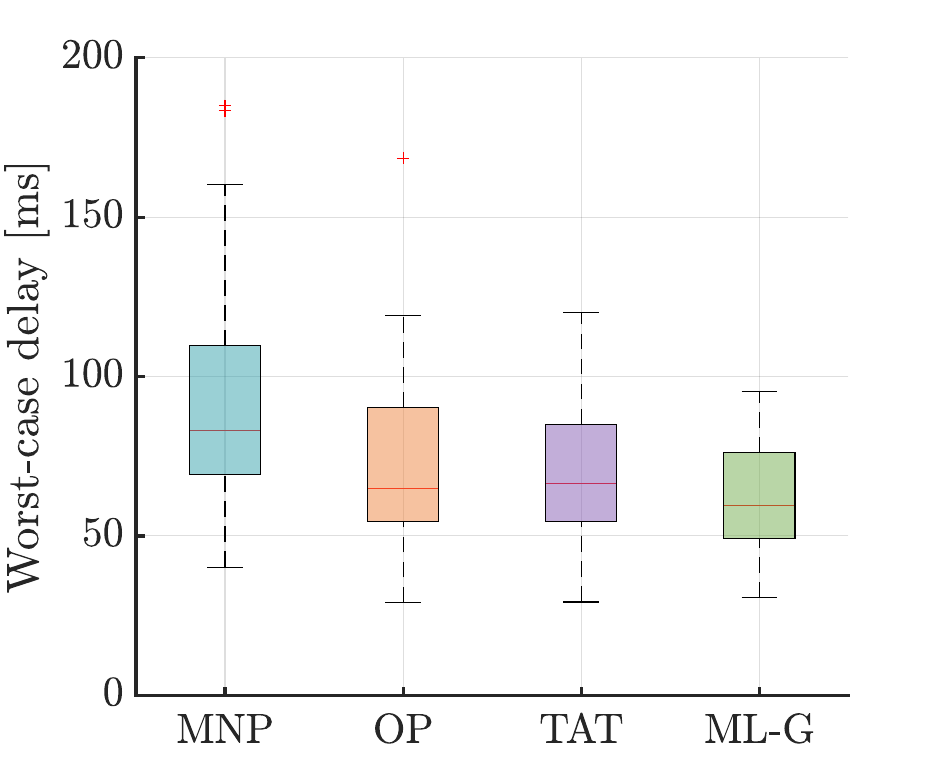}
        \caption{$N = 20$~\glspl{sta}}
        \label{fig:scalability_20stas}
    \end{subfigure}

    \caption{Worst-case delay distribution for a different number of users $N \in \{8, 12, 16, 20\}$ across 100 random deployment realizations in each scenario.}
    \label{fig:scalability_boxplot}
\end{figure*}

Fig.~\ref{fig:scalability_boxplot} shows the distribution of worst-case delays for the four scheduling strategies---\gls{mnp},~\gls{op},~\gls{tat}, and ML-G---as the number of users increases, with $J = 4$ APs and $N \in \{8, 12, 16, 20\}$ users. Each boxplot aggregates the results from 100 random deployment realizations, with a portion of the deployments discarded due to overload (ranging from 4\% for 8~\glspl{sta} to 24\% for 20~\glspl{sta}).
The total network load is kept constant (on average) across all scenarios and deployment realizations, with a mean network load of $\omega_{\text{net}} = 800$ Mb/s and a standard deviation of $\sigma_{\text{net}} = 92.4$ Mb/s. Note that this setup yields a per-user traffic distribution comparable to the previously analyzed case with 16~\glspl{sta}, i.e., $\omega_i \in [10, 90]$ Mb/s. To match each configuration, four different ML-G agents were trained using the corresponding network load distributions. 

Across different number of STAs, the ML-G agents consistently achieve a lower worst-case delay compared to the heuristic baselines. The performance gap becomes more pronounced as the number of~\glspl{sta} increases, reflecting ML-G’s better adaptability to high-density scenarios. While~\gls{tat} shows competitive behavior for $N \in \{8,12,16\}$, its performance deteriorates under scenarios with a higher number of users ($N=20$). In contrast, ML-G maintains a tighter delay distribution under 100 ms (including outliers) in all evaluated scenarios, demonstrating robust queue management and scheduling decisions under varying levels of network density, and confirming that ML-based approaches scale well in terms of delay performance as user density increases.

\section{Conclusions}
\label{sec:conclusions}

This work investigates the use of~\gls{drl} to address the~\gls{mapc} scheduling problem in IEEE 802.11bn Wi-Fi networks. We design and implement a~\gls{drl}-based agent capable of observing per-station worst delay, queue size, and channel conditions, and making scheduling decisions aimed at minimizing worst-case delay. The agent is trained using PPO within a Gymnasium-compatible Wi-Fi environment that models~\gls{csr} and heterogeneous traffic.

Our results show that the proposed model not only generalizes across diverse deployments and traffic patterns but also consistently outperforms three established~\gls{mapc} scheduling heuristics (\gls{mnp},~\gls{op}, and~\gls{tat}). 
These findings highlight the potential of learning-based approaches to improve delay-sensitive performance in coordinated Wi-Fi networks, paving the way for more intelligent scheduling in next-generation wireless standards.

Advanced scheduling strategies---such as those involving group management and transmit power control---are left for future work, along with the integration of emerging features like multi-link operation (MLO). Additionally, developing a generalizable model capable of scaling to arbitrary network sizes, regardless of the number of~\glspl{ap} and~\glspl{sta}, remains an open research direction.

\section{Acknowledgments}

This paper is supported by the CHIST-ERA Wireless AI 2022 call MLDR project (ANR-23-CHR4-0005), partially funded by AEI and NCN under projects PCI2023-145958-2 and 2023/05/Y/ST7/00004, respectively. 
The work of D. Nunez, F. Wilhelmi and B. Bellalta is also partially supported by Wi-XR PID2021-123995NB-I00 (MCIU/AEI/FEDER,UE), by MCIN/AEI under the Maria de Maeztu Units of Excellence Programme (CEX2021-001195-M), and ICREA Academia 00077.

\bibliographystyle{IEEEtran}
\bibliography{main}
\end{document}